\documentclass[journal]{IEEEtran}

\usepackage{cite}      % Written by Donald Arseneau
\usepackage{graphicx}
\usepackage{amsmath}
\usepackage{amsfonts}
\usepackage{url}
\usepackage{color}
\usepackage[normalem]{ulem}

\usepackage{mathabx}

\newcommand{\be}{\begin{equation}}
\newcommand{\ee}{\end{equation}}
\newcommand{\bea}{\begin{eqnarray}}
\newcommand{\eea}{\end{eqnarray}}
\newcommand{\beaa}{\begin{eqnarray*}}
\newcommand{\eeaa}{\end{eqnarray*}}
\newcommand{\ben}{\begin{enumerate}}
\newcommand{\een}{\end{enumerate}}
\newcommand{\bi}{\begin{itemize}}
\newcommand{\ei}{\end{itemize}}

\newcommand{\df}{{\rm d}}

\newcommand{\Eq}[1]{Eq. (\ref{#1})}

\begin{document}

\title{Efficient and accurate modeling of \\ multi-wavelength propagation in SOAs: \\ a generalized coupled-mode approach}

%\author{Cristian Antonelli, Antonio Mecozzi,~\IEEEmembership{ Fellow,~OSA, Fellow,~IEEE}\\ Mark Shtaif,~\IEEEmembership{Fellow,~OSA}, and Peter J. Winzer, ~\IEEEmembership{ Fellow,~OSA, Fellow,~IEEE}
%\thanks{Manuscript received \today}% <-this % stops a space
%\thanks{A. Mecozzi and C. Antonelli are with the Department of Physical and Chemical Sciences,
%University of L'Aquila, L'Aquila 67100, Italy., M. Shtaif is with the Department of Physical Electronics,
%Tel Aviv University, Tel Aviv 69978, Israel, P.J. Winzer is with Bell Labs, Alcatel-Lucent, 791 Holmdel-Keyport Rd., Holmdel, New Jersey 07733, USA. }}

\author{Cristian Antonelli, Antonio Mecozzi,~\IEEEmembership{Fellow,~OSA, Fellow,~IEEE}, Wangzhe Li, and Larry Coldren,~\IEEEmembership{Fellow,~OSA, Fellow,~IEEE} 
\thanks{Manuscript received \today}% <-this % stops a space
\thanks{C. Antonelli and A. Mecozzi are with the Department of Physical and Chemical Sciences,
University of L'Aquila, L'Aquila 67100, Italy; Wangzhe Li and Larry Coldren are with the Electrical and Computer Engineering and Materials Depts., University of California, Santa Barbara, CA 93106  }}

\markboth{Journal of Lightwave Technology}{Shell
\MakeLowercase{\textit{et al.}}: Bare Demo of IEEEtran.cls for
Journals}

\maketitle

\begin{abstract}

We present a model for multi-wavelength mixing in semiconductor optical amplifiers (SOAs) based on coupled-mode equations. The proposed model applies to all kinds of SOA structures, takes into account the longitudinal dependence of carrier density caused by saturation, it accommodates an arbitrary functional dependencies of the material gain and carrier recombination rate on the local value of carrier density, and is computationally more efficient by orders of magnitude as compared with the standard full model based on space-time equations. We apply the coupled-mode equations model to a recently demonstrated phase-sensitive amplifier based on an integrated SOA and prove its results to be consistent with the experimental data. The accuracy of the proposed model is certified by means of a meticulous comparison with the results obtained by integrating the space-time equations.

%We rigorously derive the coupled mode equations describing multi-wavelength mixing in semiconductor optical amplifiers (SOAs). We check the accuracy of the equations showing that the result of their solution is in excellent agreement with the results obtained by running a full time-domain simulation. We apply the obtained coupled mode equations to describe an SOA based phase sensitive amplifier, and compare the results with the data of an experiment performed on an integrated SOA.

\end{abstract}


\section{Introduction}

Semiconductor optical amplifiers (SOAs) have been in the spotlight for many years, attracting ever growing interest in multiple areas of applications. These include all-optical signal processing in fiber-optic communication networks, cost-effective local area transmission, and, more recently, integrated silicon photonics, where SOAs are the building blocks for the implementation of large-scale integrated photonic circuits. Many of these applications rely on the mixing of the wavelength components of the propagating electric field, and their theoretical study can be performed by numerically integrating the coupled nonlinear equations describing the evolution of the electric field envelope in the longitudinal direction along the SOA, and the temporal carrier dynamics \cite{connelly,Totovic}. Obviously, this approach is not suitable for the efficient design of an SOA, owing to the intensive computational effort that it involves. The search for computationally efficient and analytically tractable models has yield the formulation of what is sometimes referred to as a \textit{reduced model} for the nonlinear SOA response \cite{Agrawal}, where the space-time equations reduce to a single ordinary differential equation \cite{Agrawal}, suitable for the analytic study of multi-wave mixing (see e.g. \cite{Agrawal,MecozziOL,NLGS}). The formulation of a reduced model hinges upon two major assumptions. The first is that the spontaneous carrier recombination rate is proportional to the carrier spatial density, and the second is that the material gain also depends linearly on the carrier density. These assumptions emanate from early studies of semiconductor lasers. Indeed, in lasers the carrier density dynamics is characterized by small deviations from a steady state value which is set by the threshold condition of gain equalling the cavity loss. The small deviations around this value are only caused by amplified spontaneous emission (ASE) and by some spatial hole burning, which is however of little significance because in most structures the intra-cavity optical intensity is only moderately inhomogeneous. Consequently, in laser structures, gain and spontaneous emission rate can be accurately described by a linearized expression around the steady state carrier density. Early studies on SOA structures also used linear expressions for gain and carrier recombination, and in this case the linearization, albeit less accurate, found its ground on its simplicity and, more importantly, on the limited gain of legacy SOAs, which implied a limited longitudinal inhomogeneity of the optical field in the optical waveguide.

Unfortunately, these assumptions do not reflect the characteristics of modern SOAs, as is clarified in what follows.  Modern SOAs may have linear gain in excess to 40dB, implying a pronounced longitudinal inhomogeneity of the field intensity and hence of gain saturation. This may cause, in some cases, that the gain is only slightly saturated at the waveguide input, whereas it is almost zero at the waveguide output, where saturation is so high that the carrier density approaches its transparency value. When this is the case, a linear expression for the gain is reasonably accurate only if the gain does not deviate significantly from the linear expansion around the transparency carrier density over a range of values.
%deviation of the gain from linearity is moderate over a range of values from the transparency carrier density to the typical operating points of such high gain devices.
The nowadays widely accepted forms for the dependence of the material gain on carrier density do not meet this requirement, because  over such wide range of carrier density values the nonlinearity cannot be neglected, especially in quantum-well (QW) SOAs devices \cite{Book}. This makes the use of linear forms for the gain not an option for an accurate and quantitative description of the SOA dynamics.  In addition, advances in material fabrication have made in modern devices the contribution of defect-induced carrier recombination, which is proportional to the carrier density $N$, negligible, with the consequence that spontaneous carrier recombination is dominated primarily by radiative recombination, whose rate is proportional to $N^2$, and secondarily by Auger recombination, whose rate is proportional to $N^3$ \cite{Book}.  This reality makes the linearization of the spontaneous recombination rate also a questionable approximation. All these arguments together suggest that the accuracy of models of the nonlinear SOA response based on linearization of the carrier recombination rate and gain may be, in state-of-the-art devices, highly inaccurate.
%In spite of the growing importance of semiconductor optical amplifiers (SOAs) as building blocks of integrated photonics platforms, a set of coupled mode equations quantitatively describing multi-wavelength mixing in modern SOAs has not been derived yet. In early papers (see e.g. \cite{Agrawal,MecozziOL,NLGS}), wave mixing in SOAs was modeled assuming a linear dependence of the spontaneous carrier recombination rate and of the optical gain on carrier density $N$.
%On the one hand, it has been shown that spontaneous carrier recombination is dominated primarily by radiative recombination, whose rate is proportional to $N^2$, and secondarily by Auger recombination, whose rate is proportional to $N^3$ \cite{Book}. On the other hand, the dependence of the material optical gain on  carrier density is nonlinear, the nonlinearity being more pronounced in quantum-well (QW) SOAs rather than in bulk SOAs \cite{Book}. These nonlinear dependencies prevent the development of reduced models for the nonlinear SOA response.

A natural approach to the study of wave mixing in SOAs, which closely reminds coupled-mode theories, is the one based on the derivation of evolution equations for the complex amplitudes of the field frequency components.  Somewhat surprisingly, studies of wave mixing in modern SOAs (that is, SOAs characterized by a nonlinear dependence of the recombination rate and material gain on carrier density) based on this approach seem to be absent in the literature. In a couple of recent papers \cite{Agrawal2012,Tucker}, the authors assume a linear gain and a polynomial recombination rate, as it would be appropriate for bulk SOAs. However, they express the recombination rate as $R(N) = N/\tau_c(N)$, where $\tau_c(N) = N/R(N)$ has the meaning of an equivalent spontaneous carrier lifetime and, in the derivation of the coupled-mode equations,  they replace $\tau_c(N)$ with some time- and space-independent value. This makes, again, the assumed carrier recombination rate linear.

Another distinctive assumption of all existing coupled-mode approaches to multi-wave mixing in SOAs is that the carrier density modulation induced by the mixing is characterized by a single harmonic component \cite{Tucker}. This is a reasonable assumption when a single frequency component is dominant over the others, like for instance, in four-wave mixing (FWM) experiments where a single pump and a frequency-detuned weak signal are injected into the SOA. On the contrary, this assumption is not satisfied when multiple frequency components, detuned by a few gigahertz, have comparable intensities. This configuration characterizes for instance experiments where two strong pumps are injected at frequencies $-\Omega+\omega_0$ and $\Omega + \omega_0$, and one is interested in the amplification of a weak signal injected at the central frequency $\omega_0$. In this case, the strongest carrier modulation occurs at the beat frequency $2 \Omega$ between the two strong pumps, but the signal amplification is mainly affected by the, possibly weaker, carrier modulation at frequency $\Omega$. This configuration recently became of great interest because it describes the operation of a relevant class of SOA-based phase sensitive amplifiers (PSAs) \cite{EllisPTL,Ellis,Coldren,Coldren1}.

In this paper, we derive coupled-mode equations describing multi-wavelength mixing in SOAs characterized by arbitrary functional dependencies of the recombination rate and material gain on carrier density. These include both QW and bulk SOAs. The proposed model, which in what follows we refer to as the \textit{couple-mode model}, takes into account the frequency dependence of the material gain, as well as all orders of the waveguide dispersion, and accommodates input optical waveforms consisting of arbitrary combinations of multiple frequency components.\footnote{We consider here only the nonlinearity that comes from carrier modulation, neglecting ultrafast nonlinearity arising from carrier heating, two photon absorption and spectral hole burning. This choice has been motivated to keep the analysis simple, and also because we are interested to cases where nonlinearity is large enough to be used in all-optical processing applications or to be an issue in applications where linearity is sought for. In these cases, the frequency detuning does not exceed a few tens of gigahertz, and in this detuning range the nonlinear modulation is mostly caused by carriers. The inclusion of ultrafast processes, however, does not pose any conceptual difficulties, and can be done along the lines of ref. \cite{MecozziMork} assuming that the gain depends on quantities other than carrier density, like e.g. the carrier temperature for carrier heating, or the energy-resolved population of carriers for spectral hole burning, and assuming linear decay process of these quantities towards their steady state values.} The implementation of the model is illustrated in detail in the case of a QW SOA characterized by a logarithmic dependence of the optical gain on the carrier density $N$, and by a cubic-polynomial carrier recombination rate $R(N)$. The accuracy of the coupled-mode model is successfully tested (unlike in previous related studies) by means of a meticulous comparison with the results obtained by integrating the space-time equations of the SOA full model. Remarkably, owing to their inherent simplicity, the coupled-mode equations imply computational costs by orders of magnitude smaller that those required by the space-time equations, thus enabling the efficient characterization of multi-wave mixing in SOA structures, which would be otherwise highly impractical. We then apply the derived coupled-mode equations to studying the operation of a recently demonstrated dual-pumped PSA based on an integrated QW SOA \cite{Coldren}. We prove the results to be consistent with the experimental data, and confirm the excellent agreement with the results obtained by using the full SOA model.

\section{Coupled-mode equations for multi-wavelength propagation in SOAs}
We denote by $E(z,t)$ the slowly-varying complex envelope of the electric field propagating in the SOA in the temporal reference frame that accommodates the field group velocity $v_g$, corresponding to the real field
\be \mathcal E(z,t) = \mathrm{Re} \left[ E\left(z,t - \frac{z}{v_g} \right) e^{-i [\omega_0 t - \beta(\omega_0) z]} \right], \label{realE}\ee
with $\omega_0$ being the optical frequency. The field envelope $E$ is normalized so that that $|E|^2$ is the optical power flowing through the transverse waveguide section. It is related to the photon flux $P$ in photons per unit time and area through the relation
\be |E|^2 = \hbar \omega_0 S_{\mathrm{mod}} P, \ee
where $S_{\mathrm{mod}} = S/\Gamma$ is the modal area of the waveguide, with $S$ denoting the effective SOA area and $\Gamma$ the optical confinement factor. The evolution of $E(z,t)$ along the SOA is governed by the familiar equation
\bea \frac{\partial E}{\partial z} = \frac{1}{2} \left[(1-i\alpha) \Gamma \hat g - \alpha_\mathrm{int} \right] E + i \hat \beta  E + r_\mathrm{sp}, \label{10} \eea
where $\alpha$ is the Henry factor, $\alpha_{\mathrm{int}}$ is the SOA internal loss coefficient, and $r_\mathrm{sp}$ is the spontaneous emission noise term. By $\hat g$  and $\hat \beta$ we denote the material gain operator and the wavenumber operator. The operator formalism allows us to conveniently accommodate the frequency dependence of the gain as well as the waveguide dispersion to any order. Within this formalism the two operators can be expressed as
\bea \hat g &=& \sum_{m=0}^{\infty} \frac 1 {n!} \frac{\partial^n g (N, \omega_0)}{\partial \omega_0^n}  \left(i \frac{\partial}{\partial t}\right)^n \label{opg}\\ \hat \beta &=& \sum_{m=2}^{\infty} \frac 1 {n!} \frac{\df^n \beta (\omega_0)}{\df \omega_0^n}  \left(i \frac{\partial}{\partial t}\right)^n, \label{opn} \eea
where $g(N,\omega)$ is the gain coefficient expressed as a function of the carrier density $N$ and the optical frequency $\omega$, and $\beta(\omega)$ is the frequency-dependent field propagation constant. The expressions for $\hat g$ and $\hat \beta$ in Eqs. (\ref{opg}) and (\ref{opn}) are obtained by expanding $g(N,\omega)$ and $\beta(\omega)$ around the carrier frequency $\omega_0$. The fact that the sum in Eq. (\ref{opn}) starts from $n = 2$ is consistent with the definition of $E(z,t)$ in Eq. (\ref{realE}), which already accounts for the effect of $\beta(\omega_0)$ and $\df \beta(\omega_0)/\df \omega_0 = 1/v_g$.

The spontaneous emission noise term $r_\mathrm{sp}$ is modeled as a zero-mean, complex phase independent random process. It depends explicitly on the carrier density, besides time and space, i.e. $r_\mathrm{sp} = r_\mathrm{sp}(N,t;z)$. Its correlation function is
\be \mathbb{E} \left[ r^*_\mathrm{sp}(N,t;z) r_\mathrm{sp}(N,t';z') \right] =   \hbar \omega_0 R'_\mathrm{sp}(N,t-t') \delta(z-z') ,  \ee
where by the symbol $\mathbb{E}$ we denote ensemble averaging. Here the term $\delta(z-z')$ accounts for the fact that different longitudinal waveguide sections provide statistically independent contributions to the noise term, and
\be R'_\mathrm{sp}(N,t'-t) = \int e^{-i (\omega-\omega_0)(t'-t)} n_\mathrm{sp} (N, \omega) \Gamma g(N,\omega) \frac{\df \omega}{2 \pi}, \ee
is the spontaneous emission rate into the waveguide mode and in the field propagation direction, with $n_\mathrm{sp}$ denoting the population inversion factor. Spontaneous emission is a small perturbation of the propagating field, so that we may safely replace $N$ with its temporal average, thus neglecting the effect of its small fluctuations around this value. Within this approximation the process of spontaneous emission can be modeled as a stationary process in time.

The equation for the carrier density is
\be \frac{\partial N}{\partial t} = R_J - R_\mathrm{sp} - R_\mathrm{nr}  - R_\mathrm{st} \label{dNdt}\ee
where
\be R_J = \frac{J w_a L}{V} = \frac{J}{ed}, \ee
is the carrier injection rate into the active volume $V = SL = w_a d L$, where $w_a$ is the active region width, $L$ is the active region length, and $d$ is the active region thickness. The term $R_\mathrm{sp}(N)$ is the recombination rate associated to spontaneous emission, and its well approximated by the quadratic expression  $R_\mathrm{sp}(N) = BN^2$. By $R_\mathrm{nr}(N)$ we denote the non-radiative recombination rate, which we express as $R_\mathrm{nr}(N) = AN + CN^3$, where the linear contribution $AN$ is mostly due to defect-induced recombination, and the cubic contribution $CN^3$ to Auger recombination\footnote{We note that, while the resulting cubic polynomial expression $AN + BN^2 + CN^3$ has been shown to fit very well the experimental data in most cases \cite{Book}, the one-to-one correspondence between the three terms of the polynomial and the three recombination mechanisms is not always as neat as is illustrated in the main text. For instance, in the case of non-parabolic bands (the normal case), radiative recombination is also non-parabolic and is best modeled with a bit of linear component; carrier leakage (due to finite QW barriers) has an exponential dependence and requires a polynomial fit, affecting the numerical values of $A$, $B$, and $C$.}.
%%
%\be R_\mathrm{sp}(N) = BN^2, \ee
%%
%is the recombination rate associated to spontaneous emission, and
%%
%\be R_\mathrm{nr}(N) = AN + CN^3, \ee
%%
%is the non-radiative recombination rate, the sum of the linear (defect) recombination rate $AN$, and the Auger recombination rate $CN^3$.
As for the stimulated recombination, under the assumption that each stimulated emission process corresponds to the emission of one photon and the annihilation of one carrier, the stimulated recombination rate $R_\mathrm{st}$ (which also accounts for spontaneous emission within the waveguide mode) is related to the photon flux $P$ through
%
%\be R_\mathrm{st} = \frac{v_g}{V} \frac{\partial (V_\mathrm{mod}P) }{\partial z} = \frac{v_g}{\Gamma} \frac{\partial P}{\partial z}, \ee
\be R_\mathrm{st} S \df z = S_\mathrm{mod} \left[P(z+\df z) - P(z) \right] \Longrightarrow R_\mathrm{st} = \frac{1}{\Gamma} \frac{\partial P}{\partial z}. \ee
%
%where $V_\mathrm{mod} = V/\Gamma$ is the modal volume.
By combining the various mechanisms, Eq. (\ref{dNdt}) becomes
\be \frac{\partial N}{\partial t} = - R(N)  + \frac{J}{ed} - \frac{1}{\hbar \omega_0 S} \frac{\partial |E|^2}{\partial z}, \label{EqN1a}\ee
where by $R(N)$ we denote the familiar recombination rate\footnote{This expression of $R(N)$ is widely established and is given here for consistency with previous studies. We stress, however, that the analysis that follows does not make use of it explicitly, and rather applies to arbitrary expressions of $R(N)$.}
\be R(N) = AN + BN^2 + CN^3. \ee
By expanding the derivative $\partial |E|^2 / \partial z$ and using Eq. (\ref{10}), Eq. (\ref{EqN1a}) assumes the form
\be \frac{\partial N}{\partial t} = - R(N)  + \frac{J}{ed} - \frac{1}{\hbar \omega_0 S} \mathrm{Re} \, \left[E^* (1-i\alpha) \Gamma \hat g E\right], \label{EqN2}\ee
where we used the fact that $\hat \beta$ is a Hermitian operator, and hence it does not contribute to $\partial |E|^2 / \partial z$. The last term at the right-hand side of Eq. (\ref{EqN2}) reduces to the familiar form $\Gamma g |E|^2 / \hbar \omega_0 S$ if the gain coefficient is assumed to be frequency independent. In Eq. (\ref{EqN2}), we neglected the rate of carrier depletion associated to the photons spontaneously emitted in the guided mode, i.e. the term $R'_\mathrm{sp}(N,0)$ that comes from $\partial |E|^2/\partial z$ of Eq. (\ref{EqN1a}), because this term is small compared the carrier depletion rate caused by the photons spontaneously emitted over all spatial modes $BN^2$, which is part of $R(N)$.

We note that Eqs. (\ref{10}) and (\ref{EqN2}) can be generalized so as to include the field polarization in the analysis. While this task is rather straightforward and does not involve any conceptual challenge, we intentionally ignore polarization-related issues in order to keep the focus on the main objective of this work, which is the study of multi-wavelength propagation.

We express the multi-wavelength electric field and the carrier density as follows,
\bea E(z,t) &=& \sum_{k} E_k(z) e^{-ik\Omega t} \label{30} \\ N(z,t) &=& N_0(z) +  \sum_{k} \Delta N_k(z) e^{-ik\Omega t}, \label{40} \eea
where the coefficients $\Delta N_{k}$ must satisfy the equality $\Delta N_{-k}(z) = \Delta N_k^*(z)$ for $N(z,t)$ to be real. The term $N_0(z) + \Delta N_0 (z)$ is the $z$-dependent time-independent value of the carrier density that characterizes the system when it achieves its stationary state, and $N_0(z)$ is defined as the solution of
\be \frac{J}{e d} = R(N_0) + \frac{\Gamma}{\hbar \omega_0 S} \sum_k g(N_0,\omega_k)|E_k|^2. \label{EqN0}\ee
For values of the frequency spacing $\Omega/2\pi$ that exceed the SOA modulation bandwidth, the temporal fluctuations of $N(z,t)$ around its stationary value are filtered by the carrier dynamics and hence they can be treated within a perturbation approach. A consequence of this situation is that the deviation $\Delta N_0(z)$ of the stationary carrier density value from $N_0(z)$ is also a perturbation, and is small compared to $N_0(z)$. In this framework, all carrier-density dependent quantities that appear in Eqs. (\ref{10}) and (\ref{EqN2}) can thus be expanded to first order with respect to $\Delta N = N - N_0$, namely
\bea \hspace{-0.5cm}&&\hspace{-0.5cm} R(N) =  R(N_0) + \frac{\Delta N}{\tau(N_0)} \label{50} \\  \hspace{-0.5cm}&&\hspace{-0.5cm} g(N,\omega) = g(N_0, \omega) + g_N(N_0, \omega)\Delta N
%\equiv  g_0(\omega) + g_N(\omega) \Delta N,
, \,\,\label{60} \eea
where by the subscript $N$ we denote differentiation with respect to $N$. The quantity
\be \tau(N_0) = R_N(N_0)^{-1} = \left[ \left. \frac{\df R(N)}{\df N}\right|_{N = N_0} \right]^{-1}, \label{lifetime} \ee
is the spontaneous carrier lifetime, and
\be
%g_N(\omega) =
g_N(N_0,\omega) = \left. \frac{\partial g(N,\omega)}{\partial N}\right|_{N = N_0}, \ee
is the differential gain. We stress that \textit{these are $z$-dependent quantities}, owing to the fact that $N_0 = N_0(z)$, and hence their values evolve along the SOA. We also notice that the effective carrier lifetime governing the dynamics of carrier modulation around the steady state value is the \textit{differential} carrier lifetime $\tau(N_0)$ given in Eq. (\ref{lifetime}) and also introduced in \cite{Book53}, and not the \textit{total} carrier lifetime $\tau_c(N_0) = N_0 / R(N_0)$ used in Refs. \citeonline{Agrawal2012} and \citeonline{Tucker}. The difference between these two quantities is approximately a factor of $2$ when the radiative bimolecular recombination $BN^2$ is the dominant contribution to $R(N)$, or $3$ when the Auger recombination $CN^3$ is dominant. By inserting Eqs. (\ref{50}) and (\ref{60}) into \Eq{10} and \Eq{EqN2} we obtain
\bea \frac{\partial E}{\partial z} = \frac{1}{2} \left[(1-i\alpha) \Gamma (\hat g_0 + \Delta N \hat g_N) - \alpha_\mathrm{int} \right] E + i \hat \beta  E + r, \label{10b} \eea
and
\bea \frac{\partial \Delta N}{\partial t} &=& -\frac{\Delta N}{\tau(N_0)} - \left[R(N_0) - \frac{J}{ed}\right] \nonumber \\
&& - \frac{\mathrm{Re} \, \left[E^* (1-i\alpha) \Gamma \hat g_0 E\right]}{\hbar \omega_0 S} \nonumber \\ && - \Delta N  \frac{\mathrm{Re} \, \left[E^* (1-i\alpha) \Gamma \hat g_N E\right]}{\hbar \omega_0 S} ,  \label{100a} \eea
%
%or
%%
%\bea \frac{\partial \Delta N}{\partial t} = -\frac{\Delta N - \overline{\Delta N}}{\tau} - \frac{1}{\hbar \omega_0 S} \big\{ \mathrm{Re} \, \left[ E^* (1-i\alpha) \hat g_0 E \right] +  \Delta N \mathrm{Re} \, \left[E^* (1-i\alpha) \hat g_N E \right] \big\},  \label{110b} \eea
%%
where the operators $\hat g_0$ and $\hat g_N$ are defined as in \Eq{opg}, provided that $g(N,\omega)$ is replaced with $g(N_0,\omega)$ and $g_N(N_0,\omega)$, respectively.%, and where we defined
%%
%\be \overline{\Delta N} = \tau\left(\frac{J}{ed} - R_0\right).  \label{100c} \ee
%%
%The meaning of $\overline{\Delta N}$ is \ldots

The evolution equation for the electric field coefficient $E_k$ is obtained by inserting the expression of the field (\ref{30}) in \Eq{10b} and by equating the coefficient of the term $\exp(-ik\Omega t)$ at the two sides of the resulting equation. As a result, one finds
\bea \frac{\df E_k}{\df z} &=& \left[ \frac{1}{2}(1-i\alpha) \Gamma g(N_0, \omega_k) - \alpha_\mathrm{int}  + i \beta(\omega_k) \right] E_k \nonumber \\ && + \frac{1}{2}(1-i\alpha) \sum_n \Delta N_{k-n} \Gamma g_N(N_0, \omega_n) E_n + r_k, \nonumber \\ && \label{EqEn}\eea
where we used $\hat g_0 E = \sum_k g(N_0,\omega_k) E_k \exp(-ik \Omega t)$ and $\hat g_N E = \sum_k g_N(N_0,\omega_k) E_k \exp(-ik \Omega t)$, with $\omega_k = \omega_0 + k \Omega$. The noise term $r_k$ is defined by
\be r_k(N;z) = \int \df t e^{i k \Omega t} r_{\mathrm{sp}}(N,t;z), \ee
has zero mean $\langle r_k(N;z) \rangle = 0$, and its variance follows from
\bea && \langle r_k^*(N;z) r_h(N;z') \rangle =  \delta(z-z') \hbar \omega_0 \nonumber \\
&& \hspace{-0.5cm} \times \int \df t \int \df t' \exp[i \Omega (kt'-ht)] R'_\mathrm{sp}(N,t'-t) . \eea
Using the stationarity of $R'_\mathrm{sp}$, we may express the above as
\bea && \hspace{-0.2cm} \langle r_k^*(N;z) r_h(N;z') \rangle = \delta_{k,h}  \delta(z-z') \hbar \omega_0 \nonumber \\
&& \times n_\mathrm{sp} (N, \omega_0 + k \Omega) \Gamma g(N,\omega_0 + k \Omega). \eea
The terms $r_k(N;z')$, $k = 0,\, \pm 1,\, \pm 2\, \ldots $ are therefore a set of independent-phase, spatially-uncorrelated noise terms, which can be modeled as differentials of independent Wiener processes. At this point we can recast Eq. (\ref{EqEn}) in the following compact form
\be \frac{\df \vec E}{\df z} = \left[ \frac{1}{2}(1-i\alpha) \Gamma (\mathbf G + \mathbf{H} ) - \alpha_\mathrm{int} \mathbf I  + i \mathbf b \right] \vec E + \vec r, \label{EqEvec}\ee
where $\vec E$ and $\vec r$ are column vectors constructed by stacking the electric field coefficients $E_k$ and the noise projections $r_k$ one on top of another, respectively, with $E_0$ and $r_0$ occupying the central position, namely $\vec E = [\ldots \, E_2, \,E_1, \, E_0, \, E_{-1}, \, E_{-2} \, \ldots]^t$, and the same for $\vec r$ (the superscript $t$ stands for ``transposed''). The vector $\vec E$ and $\vec r$ are of course infinite-dimensional, and so are the square matrices $\mathbf G$, $\mathbf H$ and $\mathbf b$. Consistently with the definition of $\vec E$, we use positive and negative indices to identify the elements of these matrices, with the $(0,0)$ element occupying the central position. In particular, $\mathbf G$ and $\mathbf b$ are diagonal matrices whose $(k,k)$ elements are equal to $G_{k,k} = g(N_0, \omega_{k})$ and $b_{k,k} = \beta(\omega_{k}) - \beta(\omega_0) -  k \Omega \df \beta(\omega_0) /\df \omega_0$, respectively, whereas the $(k,n)$ element of  $\mathbf{H}$ is $H_{k,n} = \Delta N_{k-n} g_N(N_0, \omega_n)$. By $\mathbf I$ we denote the identity matrix (regardless of its dimensions).

We now proceed to the extraction of the carrier density coefficients $\Delta N_k$ by equating the terms proportional to $\exp(-i k \Omega t)$ at the two sides of \Eq{100a}, when the expression of $\Delta N$ in Eq. (\ref{40}) is inserted in it. After some straightforward algebra, involving the use of Eq. (\ref{EqN0}), one obtains
\be \left( 1 - i k \tau \Omega \right) \Delta N_k =  - \sum_{h} \Delta N_{h}  p_{k,h} + \mathcal N_k. \label{120b}  \ee
where
%{0}
\bea \mathcal N_k &=&  - \tau(N_0) R(N_0) (1-\delta_{k,0}) \nonumber \\
&& \times \sum_{n} \left[ \frac{(1-i \alpha) E_{n + k } E_{n}^*}{P_{\mathrm{stim}}(N_0, \omega_{n+k}) } + \frac{(1+i \alpha) E_{n + k } E_{n}^*}{P_{\mathrm{stim}}(N_0, \omega_{n}) }\right], \eea
\be p_{k,h} = \sum_{n} \left[ \frac{(1-i\alpha) E_{n + k - h} E_{n}^*}{P_{\mathrm{sat}}(N_0, \omega_{n + k - h})}  + \frac{(1+i\alpha) E_{n + k - h} E_{n}^*}{P_{\mathrm{sat}}(N_0,\omega_{n})} \right]. \label{DNeq}\ee
The quantity
\bea P_{\mathrm{sat}}(N_0, \omega) = \frac{\hbar \omega_0 S}{\tau(N_0) \Gamma g_N(N_0,\omega)}, \label{135} \eea
is the familiar saturation power, although its definition accounts for the frequency dependence of the gain coefficient explicitly, and
\be P_{\mathrm{stim}}(N_0, \omega) = R(N_0) \frac{\hbar \omega_0 S}{\Gamma g(N_0,\omega)}. \label{Pstim} \ee
is the power value above which carrier depletion is dominated by stimulated emission. We hence refer to $P_{\mathrm{stim}}$ as to \textit{stimulated} power. Equation (\ref{120b}) can be conveniently recast in the following compact form
\be (\mathbf I  - i \tau \Omega \mathbf k + \mathbf p) \Delta \vec N = \vec{\mathcal{N}} \label{EqNcompact}\ee
where the $(k,h)$ element  of the matrix $\mathbf p$ is equal to $p_{k,h}$, and $\mathbf k$ is a diagonal matrix with diagonal elements $\kappa_{k,k} = k$. The column vectors $\Delta \vec N$ and $\vec{\mathcal{N}}$ are constructed (like the field vector $\vec E$) by stacking the coefficients $\Delta N_k$ and $\mathcal N_k$ one on top of another, respectively, namely, $\Delta \vec N = [\ldots, \, \Delta N_1, \,\Delta N_0, \, \Delta N_{-1}, \,  \ldots]^t$ and $\vec{\mathcal{N}} = [\ldots, \, \mathcal N_{1}, \, 0, \, \mathcal N_{-1},  \, \ldots]^t$. The coefficients $N_0$ and $\Delta N_k$ are hence obtained for a given electric field state by solving Eqs. (\ref{EqN0}) and (\ref{EqNcompact}). These are the most general coupled-mode equations accounting for any functional dependence of the recombination rate and material optical gain on carrier density, as well as for the frequency dependence of the gain and waveguide dispersion.

\section{Implementation of the coupled-mode equation model in realistic SOA structures}
As is customarily done in most studies of practical relevance, where the waveguide dispersion and the frequency dependence of the gain coefficient have been shown to play a minor role, in this section we neglect chromatic dispersion, as well as higher-order dispersion, and assume frequency-independent gain. With this simplification the matrices $\mathbf G$, $\mathbf H$, and $\mathbf p$  become frequency-independent and assume a very convenient  form, as is shown in what follows. We also neglect the presence of spontaneous emission noise terms, whose implications on the SOA performance, chiefly on the SOA noise figure, will be the subject of future work.

The multi-wavelength propagation model introduced in the previous section involves an infinite number of coefficients $E_k$ and $\Delta N_k$, a situation that is obviously incompatible with its implementation in any numerical platform. However, as will be shown in the next section, high-order coefficients (namely $E_k$ and $\Delta N_k$ coefficients with large values of $|k|$) provide a negligible contribution to the solution of Eqs.  (\ref{EqN0}), (\ref{EqEvec}), and (\ref{EqNcompact}), and hence they can be omitted by truncating the vectors $\vec E$ and $\Delta \vec N$. The truncation of $\vec E$ and $\Delta \vec N$ requires of course that all matrices involved in Eqs. (\ref{EqEvec}) and (\ref{EqNcompact}) be also truncated accordingly. In what follows we provide explicit expressions for those matrices and discuss the procedure that allows the efficient computation of $\vec E$ and $\Delta \vec N$.

The truncation procedure of the infinite set of equations (\ref{EqEvec}) can be performed in a number of ways. One possible approach is assuming that $E_k(z) = 0$ for $|k| > M$. Here $M$ is an integer number that can be determined self consistently by checking that the integration of the equations for $M \to M+1$ yields indistinguishable results. This assumption implies $\Delta N_k(z) = 0$ for $|k| > 2M$, owing to the absence of beat terms at frequency offsets larger than $2 M \Omega$. A simpler yet equally accurate approach is to assume that the carrier density coefficients $\Delta N_k(z)$ are also zero at frequency offsets greater than $M \Omega$. Here we adopt the latter approach, within which Eqs. (\ref{30}) and (\ref{40}) specialize to
\bea E(z,t) &=& \sum_{k=-M}^M E_k(z) e^{-ik\Omega t} \label{30b} \\ N(z,t) &=& N_0(z) +  \sum_{k=-M}^{M} \Delta N_k(z) e^{-ik\Omega t}. \label{40b} \eea
Accordingly, the field vector $\vec E$ and carrier density modulation vector $\Delta \vec N$, consist of $(2M + 1)$ components. Matrices $\mathbf G$ and $\mathbf H$ in Eq.  (\ref{EqEvec}) become $(2M + 1)\times(2M + 1)$ matrices. In particular, owing to the assumption of frequency-flat gain, one can readily verify the equalities $\mathbf G = g(N_0)\mathbf I$, and $\mathbf H = g_N(N_0) \mathbf T(\Delta \vec N)$, where by $\mathbf T_{2M+1}(\Delta N_k)$ we denote a Hermitian-symmetric Toeplitz matrix \cite{Toeplitz}. Below we give the expression of $\mathbf T_{2M+1}(\Delta N_k)$ in the case $M = 2$ for illustration purposes,
\be \mathbf T_{5}(\Delta N_k) = \left[
                              \begin{array}{ccccc}
                                \Delta N_{0} & \Delta N_{1} & \Delta N_{2} & 0 & 0 \\
                                \Delta N_{1}^* & \Delta N_{0} & \Delta N_{1} & \Delta N_{2} & 0 \\
                                \Delta N_{2}^* & \Delta N_{1}^* & \Delta N_{0} & \Delta N_{1} & \Delta N_{2} \\
                                0 & \Delta N_{2}^* & \Delta N_{1}^* & \Delta N_{0} & \Delta N_{1} \\
                                0 & 0 & \Delta N_{2}^* & \Delta N_{1}^* & \Delta N_{0} \\
                              \end{array}
                            \right].
 \ee
The neglect of the waveguide dispersion yields $\mathbf b = 0$, and hence Eq. (\ref{EqEvec}) simplifies to
\be \frac{\df \vec E}{\df z} = \left[ \frac{(1-i\alpha)g(N_0) - \alpha_\mathrm{int}}{2} \mathbf I  + \mathbf T_{2M+1}(\Delta N_k) \right] \vec E, \label{40c}\ee
where $N_0$ is the solution of
\bea \frac{J}{e d} &=& R(N_0) \left[ 1 +  \frac{| \vec E|^2}{\tau P_{\mathrm{stim}}(N_0)} \right], \label{WP}\\  P_{\mathrm{stim}}(N_0) &=& R(N_0) \frac{\hbar \omega_0 S}{\Gamma g(N_0)}.  \label{81c}\eea
The expression for the carrier density modulation vector $\Delta \vec N$ simplifies to
\bea \Delta \vec N \hspace{-0.25cm}&=&\hspace{-0.25cm} - \frac{\tau R(N_0)}{P_{\mathrm{stim}}(N_0)} \left[\mathbf I  - \tau \Omega \mathbf k + \frac{\mathbf T_{2M+1}(C_k)}{P_{\mathrm{sat}}(N_0)} \right]^{-1} \hspace{-0.1cm} \vec{C} \label{EqNcompact2} \nonumber \\
&& \\ P_{\mathrm{sat}}(N_0) \hspace{-0.25cm}&=&\hspace{-0.25cm} \frac{\hbar \omega_0 S}{\tau(N_0) \Gamma g_N(N_0)} \label{Psat} \eea
where $C_k$ is the discrete autocorrelation function of $\vec E$, namely
\be C_k =  \sum_{n = - M}^{M} E_{n + k } E_{n}^* , \label{Ck}\ee
where we assume $E_n = 0$ for $|n| > M$. The expression of $\vec C$ in the case $M = 2$ is
\be \vec C = [C_2, C_1, C_0, C_1^*, C_2^*]^t, \ee
and that of $\mathbf T_5(C_k)$ is
\be \mathbf T_5(C_k) = \left[
      \begin{array}{ccccc}
        C_0 & C_1 & C_2 & C_3 & C_4  \\
        C_1^*   & C_0 & C_1 & C_2 & C_3 \\
        C_2^*   & C_1^*   & C_0 & C_1 & C_2  \\
        C_1^*  & C_2^*   & C_1^*   & C_0 & C_1 \\
        C_2^* & C_1^*  & C_2^*   & C_1^*   & C_0  \\
      \end{array}
    \right],
 \ee
where we used $C_{-k} = C_k^*$, as can be readily verified by inspecting Eq. (\ref{Ck}).

The numerical integration of the coupled-equations involves a three-step procedure for the transition from $z$ to $z + \Delta z$, given the field vector $\vec E(z)$. These are:
\begin{enumerate}
  \item Find the value of $N_0(z)$ by solving Eq. (\ref{WP});
  \item Extract the carrier density vector $\Delta \vec N(z)$ as in Eq. (\ref{EqNcompact2});
  \item Evaluate the field vector $\vec E(z+ \Delta z)$ by solving Eq. (\ref{40c}) from $z$ to $z + \Delta z$ while using the values of $N_0$ and $\Delta N_k$ obtained in steps 1 and 2, according to
      \bea \vec E(z+ \Delta z) \hspace{-0.2cm}&=&\hspace{-0.2cm}  \exp\left\{  \frac{(1-i\alpha)g[N_0(z)] - \alpha_\mathrm{int}}{2} \Delta z \right\} \nonumber \\ \hspace{-0.2cm}&&\hspace{-0.2cm} \exp\left\{\mathbf T_{2M+1}[\Delta N_k(z)] \Delta z \right\} \vec E(z) \eea
\end{enumerate}

%\left[ \frac{(1-i\alpha)g(N_0) - \alpha_\mathrm{int}}{2} \mathbf I  + \mathbf T_{2M+1}(\Delta N_k) \right] \vec E,

\subsection{Model validation}
\begin{figure}[t!]
\begin{centering}
\includegraphics[width=.85\columnwidth]{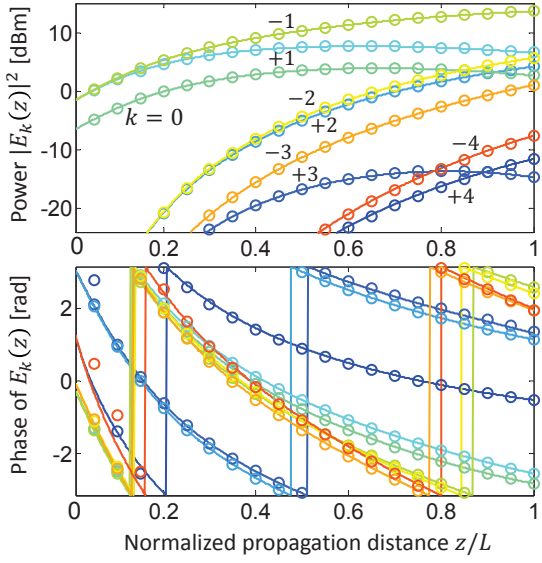}
\caption{Intensity (top panel) and phase (bottom panel) of the field components $E_k$ versus normalized propagation distance $z/L$ for the displayed values of $k$ and for the SOA parameters' values in Table \ref{Table}. Solid curves refer to the coupled-mode model, while circles were obtained by solving the space-time equations of the full model.} \label{Fig1}
\end{centering}
\end{figure}

\begin{table}[t!]
\caption{SOA parameters} % title of Table
\centering % used for centering table
\begin{tabular}{l | c | l} % centered columns (4 columns)
\hline\hline %inserts double horizontal lines
Description & Value & Units \\ [0.5ex] % inserts table
%heading
\hline % inserts single horizontal line
Linear recombination coefficient $A$ & 0 & s$^{-1}$ \\
Bimolecular recombination coefficient $B$ & $0.3\times 10^{-10}$ & cm$^3/$s \\
Auger coefficient $C$ & $3.3\times 10^{-29}$  & cm$^6/$s \\
Optical confinement factor $\Gamma$ & 10\% & \\
Linewidth enhancement factor $\alpha$ & 5 & \\
Optical wavelength $\lambda_0$ & 1561 & nm \\
Group velocity $v_g$ & $8.33\cdot 10^9$ & cm/s \\
Active region width $w_a$ & 2$\times 10^{-4}$  & cm \\
Active region tickness $d$ & 65$\times 10^{-7}$  & cm \\
Active region length $L$ & 0.1 & cm \\
Gain coefficient $g_{0}$ & 1800 & cm$^{-1}$ \\
Transparency carrier density $N_{\mathrm{tr}}$ & 2$\times 10^{18}$ & cm$^{-3}$ \\
SOA internal loss $\alpha_{\mathrm{int}}$ & 5 & cm$^{-1}$ \\
Injection current density $J$ & 3.4$\times 10^3$ & $\mathrm{A/cm^2}$\\
Frequency spacing $\Omega/2\pi$ & 8.6 & $\mathrm{GHz}$ \\ [1ex] % [1ex] adds vertical space
\hline %inserts single line
\end{tabular}
\label{Table} % is used to refer this table in the text
\end{table}
In this section we test the accuracy of the proposed multi-wavelength propagation model against the results obtained by integrating the full model' space-time equations (\ref{10}) and (\ref{EqN2}). To this end we consider a QW SOA, characterized by the following logarithmic functional dependence of the gain coefficient on carrier density \cite{Book}
\be g(N) = g_{0} \log \left( \frac{N}{N_{\mathrm{tr}}} \right), \ee
where $g_{0}$  is a gain parameter and $N_{\mathrm{tr}}$ is the carrier density required for transparency.\footnote{Of course, the use of different functional forms of $g(N)$, for instance the more accurate three parameter expression $g(N) = g_{0} \ln[(N+N_s)/(N_\mathrm{tr}+N_s)]$ also reported in \cite{Book}, is fully equivalent in terms of model complexity.} The expansion of the gain function is in this case $g(N) \simeq g(N_0) + g_N(N_0) \Delta N$, with
%
%\be g(N) \simeq g_{0} \log \left( \frac{N_0}{N_{\mathrm{tr}}} \right) + \frac{g_{0}}{N_0} \Delta N, \ee
%
%namely
%
\be g(N_0) = g_{0} \log \left( \frac{N_0}{N_{\mathrm{tr}}} \right), \hspace{0.25cm} g_N(N_0) = \frac{g_{0}}{N_0}. \ee
The physical and operational parameters of the SOA are listed in Table \ref{Table} (we note that the SOA is operated with the injection current density $J = 8.5J_{\mathrm{tr}}$, where $J_{\mathrm{tr}} = ed(A N_{\mathrm{tr}}+B N_{\mathrm{tr}}^2+C N_{\mathrm{tr}}^3)$ is the injection current density required for transparency). The SOA is injected with a three-wavelength optical signal characterized by the complex envelope
\be E_{\mathrm{in}}(t) = \sqrt{W_1} \, e^{-i \Omega t} + \sqrt{W_0} + \sqrt{W_{-1}} \, e^{i \Omega t}\ee
with $W_1 = W_{-1} = -2$dBm, and $W_0 = -7$dBm. For this set of parameters we solved the coupled-mode equations (\ref{40c}), (\ref{81c}), and (\ref{EqNcompact2}) with the input field vector $\vec E_{\mathrm{in}} = [\cdots\, 0, \, \sqrt{W_1}, \, \sqrt{W_0}, \, \sqrt{W_{-1}}, \, 0 \, \cdots  ]^t$. We used $M = 6$ and checked that larger values of $M$ yield indistinguishable results. We then integrated the full model's equations (\ref{10}) and (\ref{EqN2}) with the procedure described in \cite{Numerical}, and extracted the coefficients $E_k(z)$ from the numerical solution $E_{\mathrm{num}}(z,t)$ according  to
\be  E_{k}(z) \leftrightarrow \frac{\Omega}{2\pi} \int_{t_0}^{2\pi/\Omega} E_{\mathrm{num}}(z,t) e^{i k \Omega t} \df t , \label{Eknum}\ee
where by $t_0$ we denote any time at which the system achieved its stationary state. The results are shown in Fig. \ref{Fig1}. In the top panel we plot by solid curves the intensities of the coefficients $E_k(z)$ versus the normalized propagation distance $z/L$ for values of $k$ ranging between $k = -4$ and $k = 4$. By circles we plot the results obtained with the full model. The excellent accuracy of the coupled-mode model is self-evident. Interestingly, the figure shows that the coupled-mode model is accurate in describing the formation of four-wave mixing components that eventually (at the SOA output) exceed some of the input components. The lower panel of the same figure shows the corresponding phases of the field coefficients $E_k$ (more precisely the solid curves are the plot of $\mathrm{Phase}\left[E_k(z) \right] + k \Omega z /v_g$, where the second term accounts for the fact that the coefficients $E_k(z)$ characterize the field envelope in the time reference delayed by $z/v_g$).

\begin{figure}[t!]
\begin{centering}
\includegraphics[width=.85\columnwidth]{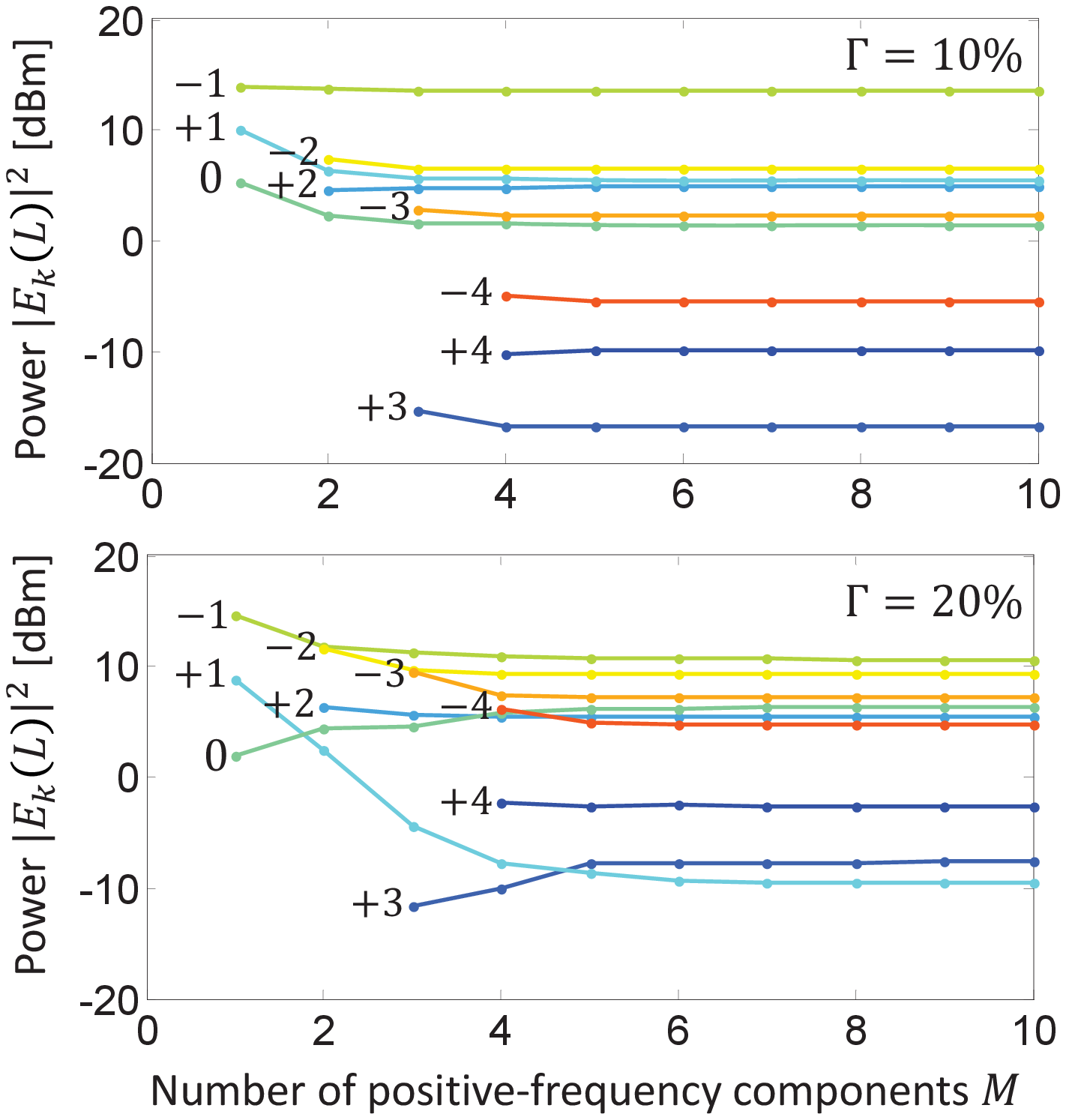}
\caption{Intensities of field components $E_k$ at the SOA output $|E_k(L)|^2$, as obtained by using the coupled-mode model, for increasing values of $M$. Each curve corresponds to a specific value of $k$ and hence it originates at $|k| = M$. The top panel refers to the set of parameters listed in Table \ref{Table} and used in Fig. \ref{Fig1}, whereas in the bottom panel the optical confinement factor was increased from 10\% to 20\%. \label{Fig2}}
\end{centering}
\end{figure}

In Fig. \ref{Fig2} we illustrate the dependence of the coupled-mode model's results on the number of field coefficients that are considered. In the top panel we plot the output intensities $|E_k(L)|^2$ evaluated by solving the coupled-mode equations for increasing values of $M$, with each curve corresponding to a different value of $k$. Since the accounting for the frequency component $E_k$ dictates that $M \ge |k|$, the curve referring to $E_k$ originates at $M = |k|$. The plot shows that in the numerical example considered here the results of the coupled-mode equations for the component $E_k$ become accurate (that is, the corresponding curve in the figure becomes flat) for $M$ exceeding $|k|$ by a one or two units. However, it should be pointed out that the convergence to the correct result is affected by the specific SOA parameters' value and may be slower. This is shown in the lower panel of the Fig. \ref{Fig2}, where the same curves plotted in the top panel are re-calculated by increasing the SOA optical confinement factor from 10\% to 20\% and by leaving the other SOA parameters unchanged. In this example, it can be seen that using $M<6$ may yield an error in the calculation of $|E_1(L)|^2$ up to a factor of 100.
%
%\lstinputlisting{Three_input_waves_versus_z_3.m}
\section{Application of the model: dual-pumped SOA-based phase-sensitive amplifier}
In this section we apply the coupled-mode model to the study of the dual-pumped SOA-based PSA presented in \cite{Coldren,Coldren1}. The goal of this exercise is two-folded. On the one hand we aim to show that the phase-sensitive gain value obtained with the coupled-mode model assuming realistic SOA parameters is consistent with the experimentally obtained value.
%As expected from the previous section, the results obtained with the coupled-mode model match those obtained with the space-time model, which implies integration times larger by orders of magnitude.
On the other hand, we show explicitly that by restricting the coupled-mode model to the pump and signal components only, as is sometimes done \cite{ChineseJLT}, yields significantly incorrect results when a realistic dependence of the amplifier gain on carrier density is used.

\begin{figure}[t!]
\begin{centering}
\includegraphics[width=.9\columnwidth]{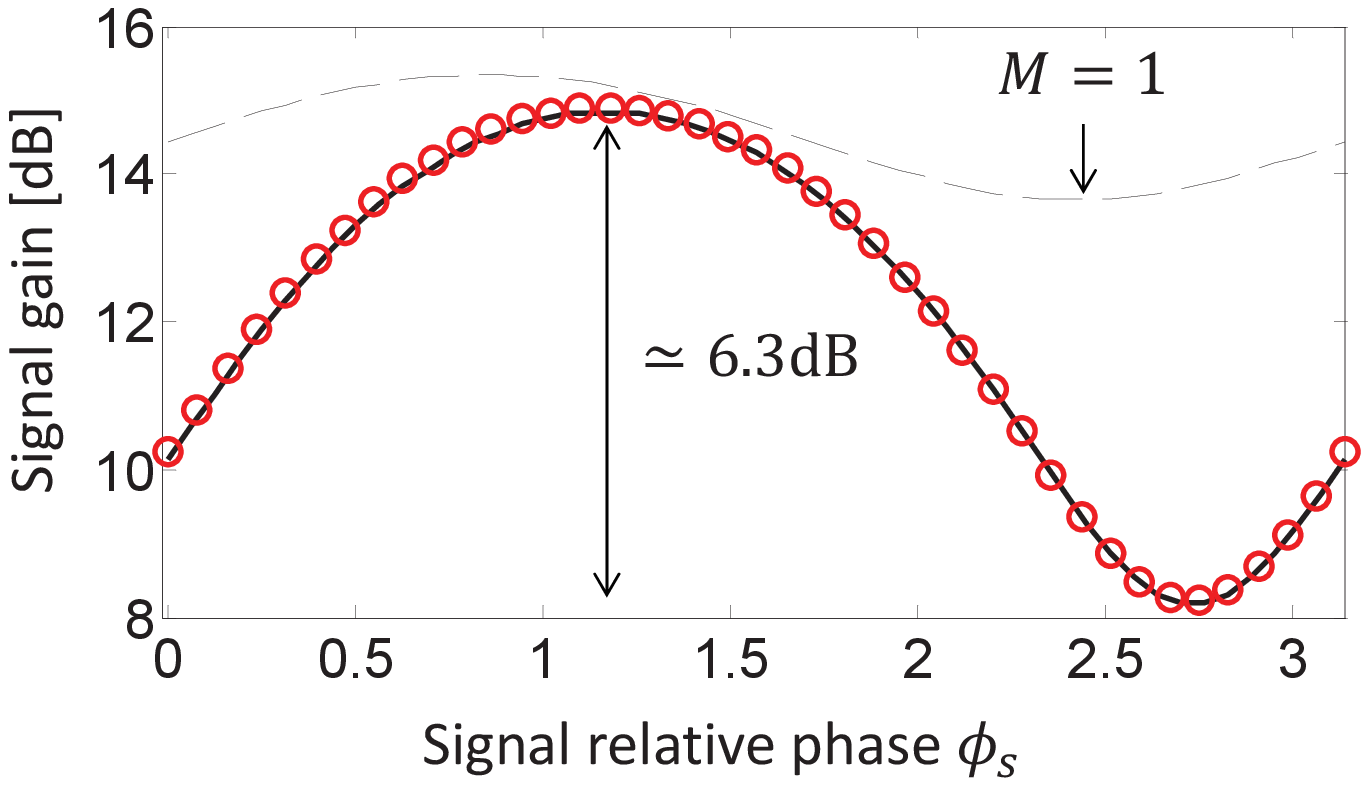}
\caption{Dual-pumped SOA-based PSA's gain versus the relative phase of the input signal $\phi_s$ introduced in Eq. (\ref{Ein}). The SOA parameters used in the numerical computation are those given in Table \ref{Table}, the input pump powers were set to $W_{P_1} = W_{P_2}=-2$dBm, and the input signal power to $W_s=-22$dBm.  The solid curve refers to the coupled-mode model with $M=4$ (larger values of $M$ yield indistinguishable results), while the circles were obtained by integrating the space-time equations of the full model. The dashed curve shows the results obtained with the coupled-mode model by propagating only the pump and signal components, namely by setting $M=1$. } \label{Fig3}
\end{centering}
\end{figure}

The waveform at the input of a PSA of the kind considered here can be expressed as
\be E_{\mathrm{in}}(t) = e^{i \phi_1}\sqrt{W_1} e^{- i\Omega t} + e^{i \phi_0} \sqrt{W_0}   + e^{i \phi_{-1}} \sqrt{W_{-1}} e^{i\Omega t}  \ee
where by $W_1$ and $W_{-1}$ we denote the optical powers of the two pumps and by $\phi_1$ and $\phi_{-1}$ their absolute phases. The field component at the central frequency represent the input signal component. By removing in all components the immaterial average phase of the two pumps $\phi_c = (\phi_1 + \phi_{-1})/2$, and denoting by $\phi_s = \phi_0 - \phi_c$ the input signal phase relative to $\phi_c$, the input field envelope can be expressed as
\be E_{\mathrm{in}}(t) = e^{i \phi_p}\sqrt{W_1} e^{- i\Omega t} + e^{i \phi_s} \sqrt{W_0}   + e^{-i \phi_{p}} \sqrt{W_{-1}} e^{i\Omega t},  \ee
where $\phi_p = (\phi_1 - \phi_{-1}) /2$. We further note that the effect of $\phi_p$ is limited to introducing an immaterial time shift $t_p = \phi_p/\Omega$, and hence it can be safely set to $\phi_p = 0$. We therefore solve the space-time equations using the following input waveform,
\be E_{\mathrm{in}}(t) = \sqrt{W_{P_1}} e^{- i\Omega t} + e^{i \phi_s} \sqrt{W_s}   + \sqrt{W_{P_2}} e^{i\Omega t},  \label{Ein} \ee
and the coupled-mode equations with the input field vector
\be \vec E_{\mathrm{in}} = [\cdots,0\, \sqrt{W_{P_1}}, \, e^{i \phi_s} \sqrt{W_s}, \, \sqrt{W_{P_2}}, 0, \, \cdots ]^t. \ee
The key quantity that characterizes the performance of the PSA under scrutiny is the dependence of the signal gain on the phase $\phi_s$. In Fig. (\ref{Fig3}) we plot the gain $G_s(\phi_s) = |E_s(L)|^2/W_s$ (in decibels) as a function of $\phi_s$, where in the case of the full model the term $E_s(L)$ is extracted from the numerical solution $E_{\mathrm{num}}(L,t)$ according to Eq. (\ref{Eknum}) with $z=L$. The SOA parameters used in the numerical example are those given in Table \ref{Table}. The input pump powers were set to $W_{P_1} = W_{P_2} = -2$dBm, and the input signal power to the much smaller value $W_{s} = -22$dBm. The solid curve is obtained by integrating the coupled-mode model with $M=4$, whereas the circles refer to the space-time model. The excellent agreement between the coupled-mode model and the space-time model, like in the previous section, is self-evident.  The thin dashed curve in Fig. \ref{Fig3} shows the result obtained with the coupled-mode model by including only the pump and signal field components, that is by using $M=1$. The plot shows that the neglect of high-order four-wave mixing products yields higher gain values and a lower phase dependent gain.

\begin{figure}[t!]
\begin{centering}
\includegraphics[width=1.0\columnwidth]{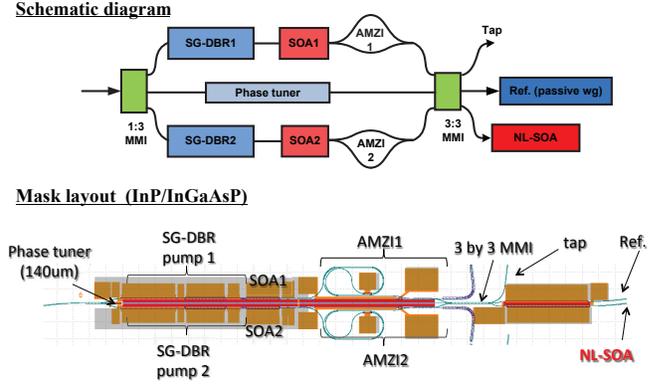}
\caption{Schematic of the signal-degenerate dual-pumped PSA and mask layout.} \label{Fig_setup}
\end{centering}
\end{figure}

Figure \ref{Fig_setup} depicts the schematic of the signal-degenerate dual pump PSA used in the experiment \cite{Coldren,Coldren1}. The coherent incident light waves, which consist of two pumps and one signal, are here generated based on external modulation and then are coupled into a PSA chip. On the chip, the input light waves are split into three paths via a 1 by 3 multimode interference (MMI) coupler. Along the upper and lower paths, there are two tunable Sampled-Grating Distributed-Bragg-Reflector (SG-DBR) lasers \cite{Coldren2}, each of which is injection locked by opposite modulation side-band pumps. Therefore, each SG-DBR laser selectively amplifies the corresponding pump and suppresses the other one as well as the signal. Signal suppression due to injection locking is necessary to avoid on-chip signal-interference-induced signal power change which otherwise could be misinterpreted as the result of PSA. After further being amplified by a downstream SOA, the pump is filtered by an asymmetric Mach-Zehnder interferometer (AMZI) to remove the residual signal and the noise falling in the signal's spectrum, which avoids signal interference among three paths and enables the signal to be shot-noise limited.

Along the middle path, there is a phase tuner to phase shift the signal based on carrier plasma effects; therefore, the adjustable and stable phase relationship among the signal and two pumps can be achieved for observing the PSA-based signal power variation as a function of the signal's phase. Please note that, although there are two pumps along the middle path, their powers are much smaller than those along the other two paths so that the pump waves along the middle path and the pump wave interference are negligible. The light waves along three paths are combined together and split again by a 3-by-3 MMI coupler to a nonlinear-SOA (NL-SOA) where phase-sensitive amplification occurs, a long passive waveguide (WG) as a reference port, and a tap to monitor the input light waves to the NL-SOA.

In the experiment, the total input power to the NL-SOA was about $-1$dBm, which was high enough to saturate the NL-SOA because the NL-SOA started saturation at $-9$dBm input power. Once the SOA was saturated, the spontaneous emission noise and PIA were suppressed. The injection current to the NL-SOA was set to be 90 mA.

\begin{figure}[t!]
\begin{centering}
\includegraphics[width=1.0\columnwidth]{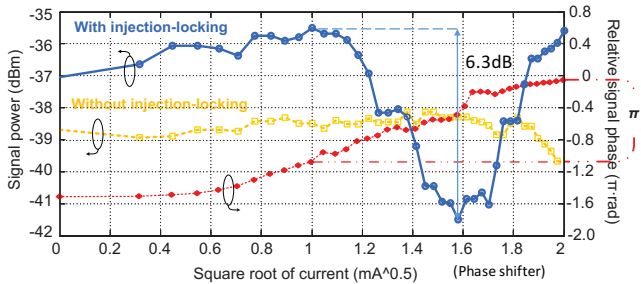}
\caption{Measured relationship among the signal power, the signal phase and the square root of the current applied to the phase tuner (see refs. \cite{Coldren1,Coldren1}).} \label{FigExp}
\end{centering}
\end{figure}

To specifically demonstrate and evaluate the PSA, the measured signal power at the output of the SOA with respect to the square root of the phase tuner current was measured, which is shown in Fig. \ref{FigExp}. The abscissa variable is set to be the square root of the phase tuner current because the signal power after PSA varies with the signal phase, which is known to vary linearly with the square root of the phase tuner current. For comparison, the signal power without injection locking and the relative phase change of the signal were measured, which are shown in Fig. 5 as well. The relative phase of the signal was obtained by using a vector network analyzer to compare the phase of the beat note between the signal and one sideband and the phase of the RF signal applied to the external intensity modulator. The beat note was generated by heterodyning the signal and only one pump wave at the output of the tap port at a photodetector. The other pump wave was suppressed by turning off the corresponding SG-DBR laser.

As can be seen from \ref{FigExp}, when injection locking was inactive and two lasers were in free-running modes, there was no PSA due to random phase drifting among the pumps and the signal waves. Once the injection locking was enabled, however, there was no obvious PSA or phase change of the signal until after the current was larger than 1 mA. Such a delay in phase shift commonly occurs in tunable SG-DBR lasers and could be caused by an N+ sheet charge that exists at the regrowth interface due to surface contamination. As the current was further increased, these traps are filled and phase-dependent signal gain appeared. Overall, one square root of the current gives one $\pi$ phase shift to the signal and one period oscillation to the signal. Clearly, such a signal power oscillation over one $\pi$ instead of $2\pi$ phase indicates that the signal power change was caused by the PSA instead of the signal interference. The measured signal power curve shows that approximate 6.3 dB extinction of phase-sensitive on-chip gain was achieved. This value is in full agreement with the results of the coupled-mode model shown in Fig. \ref{Fig3}. The linear gain that the model predicts for the simulated device is about 60dB, which is also consistent with the measured value of about 50dB \cite{Coldren1}, where the 10dB difference can be  attributed to gain compression induced by ASE and to thermal effects within the waveguide. The values of $g_0$ and of the coefficients $A$, $B$ and $C$ has been chosen as the typical parameters for the InP/InGaAsP MQW active region of the SOA used in the experiment \cite{Book}. These values have been shown to reproduce the measured static gain-current characteristic of the NL-SOA under test.

%neglect of ASE in the two models (the linear gain $G_{\mathrm{lin}}$ is the gain calculated for a vanishingly small input optical power, given by $G_{\mathrm{lin}} = \exp\left\{ [\Gamma g(N_{\mathrm{lin}}) - \alpha_{\mathrm{int}} ] L  \right\}  $, where $N_{\mathrm{lin}}$ is the solution of $R(N_{\mathrm{lin}}) = J/ed$, or equivalently of Eq. (\ref{EqN0}) with $|\vec E|^2 = 0$).

%

\section{Conclusions}

To conclude, we derived a couple-mode model for multi-wave mixing in SOAs characterized by arbitrary functional dependencies of the recombination rate and material gain on carrier density. The model takes into account the frequency dependence of the material gain, as well as all orders of the waveguide dispersion, and accommodates input fields consisting of arbitrary combinations of multiple frequency components. We showed that the conventional approach assuming a limited number of generated four-wave mixing components gives inaccurate results when two waveforms of similar intensities are injected into the SOA. In this case, our model gives highly accurate results if a sufficient number of generated components are taken into account, as we showed by direct comparison with full time-domain simulations. We applied the coupled-mode model to studying the operation of a recently demonstrated dual-pumped PSA based on an integrated QW SOA \cite{Coldren}, and showed that the outcome of the model is consistent with the experimental results.

\section*{Acknowledgement}
Work funded by DARPA through project W911NF-14-1-0249.  C. Antonelli and A. Mecozzi also acknowledge financial support from the Italian Ministry of University and Research through ROAD-NGN project (PRIN 2010-2011), and under Cipe resolution n. 135 (Dec. 21, 2012), project INnovating City Planning through Information and Communication Technologies (INCIPICT).

\end{document}